# MS$^4$: a BPHZ killer

N.D. Lenshina[1], A.A. Radionov[2], F.V. Tkachov[2]
[1] Physics Department, Lomonosov Moscow State University
[2] Institute for Nuclear Research, Russian Academy of Sciences, Moscow

lenshina.nd14@physics.msu.ru
alex.radionov89@gmail.com
fyodor.tkachov@gmail.com

**Abstract**

The UV renormalization scheme MS$^4$ emerged in the formalization of the reasoning which yielded an array of important algorithms in the 80's. MS$^4$ guarantees finiteness of renormalized integrals by construction, satisfies the Stueckelberg-Bogolyubov causality axiom for the R-operation, and turns out to be a 4-dimensional analog of t'Hooft's MS-scheme. The well-known IBP reduction algorithm can be ported to MS$^4$ with modifications, but without problems. MS$^4$ exhibits transparency of the structure, simplicity of the arithmetic at $D = 4$, and new calculational options. A straightforward derivation of RG equations runs in terms of explicitly finite quantities and expresses RG functions in terms of explicitly finite integrals.

**Introduction**

This project started as an educational one because of a lack of comprehensible treatment of the UV R-operation in the literature. The BPHZ theory [1] is esoteric and far from calculational practice within the dimensional regularization and the MS scheme [2]. On the other hand, the helicity amplitude methods for radiative corrections and the supersymmetry (cf. [3]) are ill-compatible with $D \neq 4$. Therefore, we attempted to study in practical examples the renormalization scheme introduced in [4] (we call it MS$^4$ henceforth) that has been remaining outside the scope of attention of the theoretical mainstream despite the fact that it emerged in the formalization [5] of the reasoning which yielded an array of breakthrough results (the R*-operation [6], operator expansions for models with massless particles with efficient formulas for coefficient functions [7], the asymptotic operation — Euclidean [8] and non-Euclidean [9]). However, the project surpassed the boundaries of a student exercise as it ran into surprises.

## The main dilemma of the theory of UV R-operation

The basic structural requirement for the UV R-operation comes from the causality conditions [11], [12] and requires that it must be equivalent to adding quasi-local counterterms, e.g.:

$$R_{UV}\mathrm{T}\big(\mathcal{L}(x)\mathcal{L}(y)\mathcal{L}(z)\big) = \mathrm{T}\big(\mathcal{L}(x)\mathcal{L}(y)\mathcal{L}(z)\big) + \mathrm{T}\big(\Lambda(x,y)\mathcal{L}(z)\big) \\ + \mathrm{T}\big(\Lambda(x,z)\mathcal{L}(y)\big) + \mathrm{T}\big(\Lambda(y,z)\mathcal{L}(x)\big) + \Lambda(x,y,z) \qquad (1)$$

To complete the definition, one should fix the counterterms $\Lambda$ and prove that the resulting loop integrals are finite.

The formula has a generalized-functional nature while being recursive (a T-product is defined in terms of T-products with a lesser number of $\mathcal{L}$'s). The recursive structure can theoretically be a basis for compact inductive proofs. However it can only be preserved if the techniques of generalized functions are used. The dilemma that emerges here is as follows:

A. To get rid of generalized functions in favor of ordinary integrals. For this, however, one has to resolve the recursion and forego its advantages. This is the BPHZ approach that generates a rather cumbersome formalism (cf. the cumbersome computer programs which emerge when branching recursions are resolved into embedded loops). The proof of finiteness of the resulting cumbersome integrals constitutes the purpose of the Bogolyubov-Parasyuk-Hepp theorem.

B. The other option is to fully exploit the recursion, to which end one has to develop a special generalized-functional techniques to work with products of singular functions (see below for a discussion of such mathematical tools). However, a literal implementation of this approach implies the use of coordinate representation; this line of reasoning was implemented in [13].

However, there is another option, namely, the MS$^4$ scheme that preserves the recursion and allows one to exploit its advantages, but it is formulated in the momentum representation which is the most natural one for physical problems.

## Basic mathematical tools

Two special generalized-functional constructs are used in the definition of the MS$^4$ scheme (complete details can be found in [5]):

1. The R-operation that yields a generalized function for a non-integrable product of singular functions:

$$\mathbf{R} \circ G = \mathbf{r} \circ \mathbf{R}' \circ G, \quad \mathbf{R}' \circ G = \sum_{\Gamma \triangleleft G} \big(\chi_\Gamma G \setminus \Gamma\big)\big\{\mathbf{R} \circ \Gamma\big\}, \qquad (2)$$

where the summation runs over senior divergent subgraphs (special subproducts), the functions $\chi_\Gamma$ form a "conical" decomposition of unit, (...) is a smooth factor, {...} is a singular factor. $\mathbf{r}$ is a subtraction operator for a singularity at an isolated point that is similar to one used in [14] where it is called "regularization"; such operators in quantum field theory require a further concretization to take into account the structural specifics of multidimensional perturbative integrals, for details see [13].

2. The asymptotic operation (As-operation) that yields the asymptotic expansion in a small parameter in the sense of generalized functions for a singular product:

$$\mathbf{As}\circ G = \mathbf{r}\circ \mathbf{As}'\circ G + c_G\,\delta_G, \quad \mathbf{As}'\circ G = \sum_{\Gamma \triangleleft G}\bigl(\chi_\Gamma \mathbf{T}\circ G\setminus\Gamma\bigr)\bigl\{\mathbf{As}\circ\Gamma\bigr\}, \qquad (3)$$

where $\mathbf{T}$ is a Taylor expansion in the expansion parameter. The $\mathbf{As}$ definition differs from that of $\mathbf{R}$ essentially by the fine-tuning of $\mathbf{r}$ by finite counterterms $c_G\,\delta_G$ (remember that similar counterterms $c_\Gamma\,\delta_\Gamma$ recursively emerge within $\{\mathbf{As}\circ\Gamma\}$); here $\delta_G$ are localized at the singularity point (i.e. $\delta_G$ are derivatives of the $\delta$-function) and $c_G$ are specially chosen numerical coefficients that happen to contain the non-analytical dependences on the expansion parameter [8].

The compactness of these quite non-trivial formulas (which go beyond the framework of e.g. the formalism of ref. [14]) is a consequence of their recursive structure: the task is sequentially reduced to similar subtasks for subproducts with a lesser number of factors, making possible efficient derivations and proofs by induction.

## The minimal subtraction scheme MS⁴

For the role of UV R-operation one here chooses the explicit expression for finite counterterms $c_G$ that automatically emerges in the construction (3):

$$R_{UV}\int dp\,B(p;\kappa) \triangleq \lim_{\Lambda\to\infty}\int^\Lambda dp\bigl[B(p;\kappa) - \mathbb{S}(p;\kappa)\bigr] \qquad (4)$$

The notations are as follows: $p$ is the set of loop momenta; $\kappa$ is the set of external momenta and masses; $B$ is the "bare integrand" prior to any subtractions; $\mathbb{S}$ are the so called "shadow terms" defined and discussed in what follows. $\int^\Lambda dp = \int dp\,\Theta(p/\Lambda)$ is an upper cutoff by means of a smooth function equal to one around zero and to zero around infinity. Such functions often emerge when one works with the MS⁴ scheme. The functions $\Theta(p)$ for different perturbative integrals are best chosen in a coherent fashion; this is easiest achieved by choosing them as products of elementary factors $\Theta(p_l^2)$ that correspond to propagators $D(p_l)$.

The key point is that the right-hand side of eq. (4) is finite by construction, therefore, instead of proving an analog of the BPH theorem, one only has to prove that the definition satisfies the causality axiom (1); this task turns out to be much simpler. (This effect of proof simplification due to the recursive structure as compared with the BPHZ theory is more directly seen in the coordinate space reasoning of ref. [13] to be contrasted with the coordinate space version of BPHZ proof presented in [15].)

## Shadow terms $\mathbb{S}$

These are the terms of the asymptotic expansion in the sense of generalized functions for $B$ at $p \to \infty$ that generate divergences in the integral at $\Lambda \to \infty$. They are constructed as follows:

1. The asymptotic expansion in the sense of generalized functions is done on the space without the point $p = 0$ (i.e. the expansion is valid on all test functions whose support does not contain $p = 0$):

$$B(p;\kappa) \xrightarrow{p \to \infty} \mathbf{As}' \circ B(p;\kappa); \qquad (5)$$

2. Retained are only the terms that generate divergences after integration as $\Lambda \to \infty$:

$$\mathbf{As}' \circ B(p;\kappa) \to \mathbf{As}'_0 \circ B(p;\kappa); \qquad (6)$$

3. The addenda which generate logarithmic divergences after integration as $\Lambda \to \infty$ are also logarithmically divergent near $p = 0$ (cf. the integral $\int_0^\infty dx/x$). Therefore, a further subtraction operator $\mathbf{r}$ is applied to define the generalized function on the entire space including the point $p = 0$ (i.e. to make the generalized function well-defined on any test functions without restrictions on their support):

$$\mathbf{As}'_0 \circ B(p;\kappa) \to \mathbf{r} \circ \mathbf{As}'_0 \circ B(p;\kappa) \triangleq \mathbb{S}. \qquad (7)$$

This completes the definition of the shadow terms.

### Properties of the shadow terms

First of all the expression $\mathbf{As}' \circ B$ is defined uniquely, therefore the operation $\mathbf{As}'$ commutes with linear transformations of $B$ and with its multiplications by polynomials of $p$ [5].

The operator $\mathbf{r}$ contains a finite arbitrariness which is partially (almost completely) fixed by the requirements that are specific to perturbative QFT:
1. some simple restrictions that ensure coherence between different $\mathbf{r}$'s for different perturbative integrals;
2. standard restrictions to ensure the gauge identities;
3. an exact commutation with multiplication of $B$ by polynomials of $p$ (and then one has to evaluate commutators of $\mathbf{r}$ with derivatives; this is needed e.g. for derivation of IBP identities).

The remaining arbitrariness in $\mathbf{r}$ can be parametrized by finite counterterms localized at $p = 0$. Such an arbitrariness is a familiar feature whenever the infinites are subtracted/regularized.

Now it is easy to understand the name "shadow terms": under the chosen restrictions for $\mathbf{r}$ the following property holds. If $B = \mathcal{P}_1 B_1 + \mathcal{P}_2 B_2$, where $\mathcal{P}_i$ are any polynomials of $p$, then $(B - \mathbb{S}) = \mathcal{P}_1 (B_1 - \mathbb{S}_1) + \mathcal{P}_2 (B_2 - \mathbb{S}_2)$. (Mathematically speaking, a structure of a module over a ring of polynomials [16] emerges here in a natural fashion; whether or not it may have a significance for the theory of the MS⁴ scheme, is a question that we leave to experts in pure mathematics.)

The recursion manifests itself via the fact that the coefficients $c_\Gamma$ within **As**$'$ are similar integrals but with a lesser number of factors in the integrand and a lesser dimensionality of the integration space (the familiar structure graph → subgraph).

The finiteness of the integral (4) is guaranteed by the way the shadow terms are constructed and the limit $\Lambda \to \infty$ is taken. It is easy to demonstrate that the limit value is independent of the choice of the cutoff function $\Theta(p/\Lambda)$.

## The correct structure of UV R-operation

Let us use a 2-loop example to show how the correct structure of UV R-operation (1) is restored (the adding of counterterms multiplied by integrals for graphs with UV subgraphs shrunk to points). Denote $B(p;q) = p^{-2}(p-q)^{-2}$ (one loop) and $B(p,q;k) = B(p;q)B(q;k)$ (two loops), where $k$ is an external momentum. A direct calculation using (7) and properties of **r** and $\Theta$ yields, under the limit $\Lambda \to \infty$:

$$\int^\Lambda dp\, dq\, [B(p,q;k) - \mathbb{S}] = \int^\Lambda dp\, dq\, B(p,q;k) + Z_1(\Lambda) \int^\Lambda dq\, B(q;k) + Z_2(\Lambda) + o(1), \quad (8)$$

where $Z_1(\Lambda)$ and $Z_2(\Lambda)$ are infinite counterterms, and $o(1)$ comprises terms that vanish in the limit $\Lambda \to \infty$. $B(q;k)$ corresponds to the graph with UV subgraph shrunk to a point, as required.

## Comparison with the usual MS-scheme

First, the MS⁴ scheme (or, to be precise, the family of subtraction schemes differing by the choice of finite parts in the definitions of **r**, etc.) is equivalent to the family of the usual MS [2], $\overline{\text{MS}}$ [17], G-scheme [18], etc. within $D \neq 4$, which is seen after a formal introduction of $D \neq 4$: the subtraction operator **r** corresponds to a subtraction of poles in $D - 4$, modulo the usual finite arbitrariness. Second, in the intermediate calculations, the MS⁴ scheme allows one to employ any convenient intermediate regularization, which can be effectively used. Compare the calculation of the integral of $B(p,q;k)$ (it appears e.g. in the IBP algorithm of [19]) within the MS⁴ scheme with an auxiliary analytical regularization ($B_\alpha(p,q;k) = B_\alpha(p;q)B(q;k)$, where the parameter $\alpha$ is introduced into one propagator only: $p^2 \to p^{2(1+\alpha)}$):

$$\int d^4q\, B(q;k) \int d^4p\, B_\alpha(p;q) \sim k^{-2\alpha} [\alpha(1-\alpha)]^{-2} \quad (9)$$

and the analogous calculation within the dimensional regularization (where one modifies not the integrand but the integration measure: $d^4p \to d^D p$; $D = 4 - 2\varepsilon$):

$$\int d^D q\, B(q;k) \int d^D p\, B(p;q) \sim \frac{1}{k^{4\varepsilon}} \frac{1}{2\varepsilon^2 (1-2\varepsilon)(1-3\varepsilon)} \frac{\Gamma^3(1-\varepsilon)\Gamma(1+2\varepsilon)}{\Gamma(1-3\varepsilon)} \qquad (10)$$

One can see that the arithmetic proper is rather simpler in MS⁴, and this difference is expected to naturally increase with the number of loops.

## Porting IBP algorithms to MS⁴

Consider the well-known IBP algorithms [19] (such algorithms would remain useful within MS⁴ at least for calculations of OPE coefficient functions [7], if not for RG calculations; see below). On the whole, the identities are derived similarly to the dimensionally regularized case, but there are little differences. Within MS⁴, some new identities emerge, for instance (this one is easily verified by direct calculation):

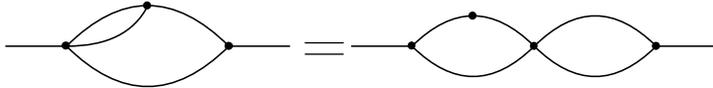

All identities become noticeably simpler (coefficients are always rational *numbers* rather than rational *functions* of $D$). On the other hand, the number of irreducible integrals is larger; but then they can be evaluated within any convenient regularization (even within the dimensional one, which, however, is not necessarily the optimal choice). For example, the following well-known integral becomes irreducible (which is of little concern in practice since one can still do it within dimensional regularization, if so desired):

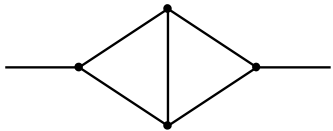

On the whole, one expects that the calculations will require somewhat more analytical preparation, but are bound to be executed much faster (a familiar trade-off). In particular, in MS⁴ one does not have to deal with the pesky interference of the dependencies on $\varepsilon = (4-D)/2$ in the (accumulating) coefficients of the identities and in the answers for irreducible integrals, which is typical for $D \neq 4$ (although some remnants of that problem will be felt at the stage of analytical preparation); such arithmetic interference is known to encumber calculations (a bug of this nature affected the first version of the $O(\alpha_s^3)$ calculation of $\sigma_{\text{tot}}(e^+ e^- \to \text{hadrons})$ [20]).

## RG equations

The differential RG equations are derived within MS⁴ in a most straightforward fashion — by a direct differentiation with respect to $\mu$ that is hidden in the definition of the subtraction operators **r**. Consider the renormalized integral (4), taking into account (5)–(7), with the dependence on $\mu$ explicitly shown:

$$A_\mu(\kappa) = \lim_{\Lambda\to\infty} \int dp\, \Theta(p/\Lambda)\left[B(p;\kappa) - \mathbf{r}_\mu \circ \mathbf{As}'_0 \circ B(p;\kappa)\right], \qquad (11)$$

The calculation exploits the fact that the bare integrand $B$ and the operation $\mathbf{As}'$ do not depend on $\mu$. Taking into account the definition of **r**, one obtains:

$$A_\mu(\kappa) = \lim_{\Lambda\to\infty}\left[\{\ldots\} + \int dp\, \Theta\!\left(\frac{p}{\mu}\right) \circ \mathbf{as}'_0 \circ B(p;\kappa)\right] \qquad (12)$$

where the curly braces contain the terms that are independent of $\mu$, whereas $\mathbf{as}'_0$ retains only those terms from $\mathbf{As}'_0$ which correspond to logarithmic divergence. Differentiating with respect to $\mu$ yields:

$$\mu\frac{\partial}{\partial\mu} A_\mu(\kappa) = \lim_{\Lambda\to\infty} \int dp\, \mu\frac{\partial}{\partial\mu}\Theta\!\left(\frac{p}{\mu}\right)\left[\mathbf{as}'_0 \circ B(p;\kappa)\right], \qquad (13)$$

Convert the derivative with respect to $\mu$ into a derivative with respect to $p$ using the Euler identity:

$$\mu\frac{\partial}{\partial\mu}\Theta\!\left(\frac{p}{\mu}\right) = (-)p^\alpha \frac{\partial}{\partial p_\alpha}\Theta\!\left(\frac{p}{\mu}\right) = (-)\hat\partial\Theta\!\left(\frac{p}{\mu}\right), \qquad (14)$$

replace $p \to \mu p$, and drop the limit with respect to $\Lambda$ that has become superfluous:

$$= -\int dp\, \hat\partial\Theta(p)\, \mathbf{as}'_0 \circ B(p;\kappa/\mu). \qquad (15)$$

One then obtains a differential RG equation for a single diagram:

$$\mu\frac{\partial}{\partial\mu} A_\mu(\kappa) + \int dp\, \hat\partial\Theta(p)\, \mathbf{as}'_0 \circ B(p;\kappa/\mu) = 0. \qquad (16)$$

Now expand the definition of $\mathbf{as}'_0$ and use the assumed properties of $\Theta(p)$:

$$\mu\frac{\partial}{\partial\mu} A_\mu(\kappa) + \sum_{\substack{\gamma=\varnothing,\ldots,\Gamma \\ \gamma\neq\Gamma}} c^*_\gamma\!\left(\frac{k}{\mu}\right) \times \left(\int dp_{\Gamma\setminus\gamma}\left[\hat\partial_{\Gamma\setminus\gamma}\Theta(p_{\Gamma\setminus\gamma})\right]\mathbf{R}'_{\mu=1} \circ \mathbf{t}_0 \circ B_{\Gamma\setminus\gamma}(p_{\Gamma\setminus\gamma};0)\right) = 0, \qquad (17)$$

where $\Gamma$ is the initial graph (the product); $\gamma$ are the complete IR singular subproducts [8]; $\Gamma\setminus\gamma = \prod_i G_i$ is a product of UV subgraphs $G_i$; $c^*_\gamma$ is an integral $\Gamma$ with all $G_i$ contracted to points (at work here is the same combinatorics which ensured that the definition (4) satisfies the causality axiom (1)); $\mathbf{t}_0$ retains only those terms of the Taylor expansion in the external parameters of the subgraph, which correspond to logarithmic divergence at $p_{\Gamma\setminus\gamma} = 0$; finally, the integral $\left(\int dp_{\Gamma\setminus\gamma}\ldots\right)$ automatically factorizes into a product of similar integrals for the UV subgraphs $G_i$.

By comparing this equation for a single graph with differential RG equations for complete perturbative sums of graphs, one realizes that the factors of this product will become — after appropriate resummations and reshufflings — contributions to RG functions ($\beta$, $\gamma$, ...) from $G_i$.

### Example

Consider a special case of equation (17) for the following 2-loop graph of the $g\varphi^4$ model with an UV subgraph (both graph and the subgraph here are logarithmically divergent):

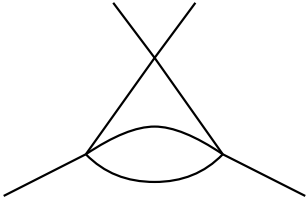

One obtains:

$$\mu \frac{\partial}{\partial \mu} A_\mu(k,k') + \lim_{\Lambda \to \infty} \int^\Lambda dq \left[ B(q;k) - \mathbb{S}_\mu \right] \times \int dp \left[ p^\alpha \frac{\partial}{\partial p^\alpha} \Theta(p) \right] B(p;0) = 0 \,. \tag{18}$$

One sees that the first integral is exactly the integral obtained by shrinking the UV subgraph to a point and renormalized according to the MS⁴ definition (4), whereas the second integral contributes to the lowest order term of the $\beta$-function of the $g\varphi^4$ model.

### Building blocks for RG functions

One concludes that after resummation of the perturbation series, RG functions will be assembled from integrals of the following form that are finite by construction:

$$\Delta_G = \int dp \, \hat{\partial}\Theta(p) \times \mathbf{R}'_{\mu=1} \circ \mathbf{t}_0 \circ B(p;0), \tag{19}$$

where $G$ is an UV subgraph; $p$ is the collection of its loop momenta; $\hat{\partial}$ is the Euler operator (14) with respect to $p$; $\mathbf{t}_0$ retains only those terms of the Taylor expansion with respect to external momenta and masses of the subgraph, which correspond to logarithmic divergence at $p=0$. $\Theta(p)$ is a smooth function that is equal to 1 (0) near 0 ($\infty$); its derivatives are zero in either case, therefore no problems arise with convergence of the integral near zero and infinity, whereas the non-integrable singularities in the subproducts are eliminated by $\mathbf{R}'$. Note that $\mathbf{R}'_{\mu=1}$ is taken at $\mu=1$, which means that $\Delta_G$ (modulo trivial overall dimensional coefficients), and therefore RG functions, are dimensionless quantities, which can be regarded as a characteristic property of the entire class of renormalization schemes MS, $\overline{\text{MS}}$, G, MS⁴, etc.).

## Conclusions

To summarize, the MS$^4$ scheme guarantees finiteness of the resulting integrals by construction and satisfies the Stueckelberg-Bogolyubov causality axiom, thus being a correct UV R-operation. It does not require intermediate regularizations on top of the natural cutoffs that are inevitably implied by the definition of improper integrals. The scheme has a clean structure: technicalities are packaged into universal tools ($\Theta$, **r**, **R**, **As**). The scheme directly connects to the mainstream calculational practices: the family of MS schemes is restored by a formal introduction of dimensional regularization. Standard IBP algorithms that are usually formulated within $D \neq 4$ can be ported to MS$^4$ mutatis mutandis. In contrast with the dimensional regularization, working within MS$^4$ requires that one fully understands a non-trivial formalism, but then the arithmetic of the calculations (e.g. IBP identities) is noticeably simpler than with $D \neq 4$, the difference increasing with the number of loops. To put it another way, the apparent (some would say dumb as in "Dumb simplicity is why the show is a big hit") simplicity of the dimensional regularization turns out to carry a hidden cost.

New and unusual explicit expressions in terms of finite integrals are obtained for RG functions; here one can expect new and unusual computational options. There are no difficulties with operator expansions etc., as the MS$^4$ scheme sprang up within the theory of asymptotic operation that was developed exactly for this kind of problems [8], [5], so MS$^4$ and OPE are a perfect match, technically.

We conclude that the MS$^4$ scheme is a full-fledged alternative to the BPHZ theory, has important theoretical and technical advantages at the level of the formalism, as well as a direct connection to applications. Finally, the MS$^4$ scheme exhibits unusual features that are of interest for calculational applications.

## Acknowledgments

The authors thank R.Sh. Menyashev for supporting their participation in the 2019 Bogolyubov Conference and D.I. Kazakov for a help with the corresponding arrangements.